\documentclass[superscriptaddress,prl,reprint,aps]{revtex4-1}
\usepackage{CJK}
\usepackage{graphicx}
\usepackage{dcolumn}
\usepackage{bm}
\usepackage{amssymb,amsmath}

\begin{document}

\title{3D Single-port Labyrinthine Acoustic Metamaterial}
\author{Chi Zhang}
\author{Xinhua Hu}
\email{huxh@fudan.edu.cn}
\affiliation{Department of Materials Science, Key Laboratory of Micro- and Nano-Photonic
Structures (Ministry of Education), and Laboratory of Advanced Materials,
Fudan University, Shanghai 200433, China}
\date{\today }

\begin{abstract}
In this paper, we report on the design, fabrication, and experimental
characterization of a 3D single-port labyrinthine acoustic metamaterial. By
using curled perforations with one end closed and with appropriate loss
inside, the proposed metamaterial can perfectly absorb airborne sounds in a
low frequency band. Both the position and width of the band can be tuned
flexibly. A tradeoff is uncovered between the relative absorption bandwidth
and thickness of the metamaterial. When the relative absorption bandwidth is
as high as $51\%$, the requirement of deep subwavelength thickness ($%
0.07\lambda $) can still be satisfied.
\end{abstract}

\pacs{43.20.+g, 43.58.+z, 46.40.Cd}
\maketitle

Acoustic metamaterials (AMMs) are artificial periodic structures with
subwavelength building blocks that exhibit unusual acoustic characteristics
\cite{Liu,Li1,Lai,Fang,Hu,Yang1,Mei,Yang2,Ma}. There has been tremendous
attention in AMMs since its first demonstration in 2000 by Liu \textit{et al}
\cite{Liu}. An amount of functionalities and applications have been proposed
and achieved based on AMMs \cite{Li2,Zhu,Park,Lu,Chr2,Fle,Cum,Chen,Zhang}.
Labyrinthine AMMs (LAMMs) composed of curled perforations are one of the
most significant types of AMMs due to their extreme constitutive parameters
and plentiful potential applications \cite{Liang1,Xie1,Liang2,Xie2,Cheng}.
For instance, LAMMs show diverse properties such as double negativity, a
density near zero, and a large refractive index in different frequencies,
giving rise to fascinating phenomena including negative refraction and
zero-density tunneling \cite{Liang1,Xie1,Liang2}. By applying a graded
structures, a labyrinthine metasurface with graded index can be constructed
to modify the wavefront and direction of outgoing waves \cite{Xie2}. Total
reflection of low-frequency sounds has also been achieved very recently by
applying an ultrasparse labyrinthine metasurface \cite{Cheng}. By now, most
of previous LAMMs have a two-port character so that transmission is
permitted and absorption can be reasonably neglected. In addition, the
studies have been mainly done in 2D cases. Although they are important to
fully control sound waves in 3D space, 3D LAMMs have seldom been
investigated and demonstrated.

In this paper, we report on the design, fabrication, and experimental
characterization of 3D single-port LAMMs that are composed of curled,
one-end-closed channels. Via adjusting the sound loss in channels to a
critical value, such LAMMs can have impedance matching to background and
thus perfectly absorb sounds in a low frequency band; both the position and
width of the band can be tuned flexibly. Analytic formulas are derived to
predict the critical loss in channels and relative absorption bandwidth, and
their accuracy are verified by both simulations and experiments. A tradeoff
is uncovered between the relative absorption bandwidth and thickness of the
LAMM. When the relative absorption bandwidth is as high as $51\%$, the
requirement of deep subwavelength thickness ($0.07\lambda $) can still be
satisfied.

\begin{figure*}[tbp]
\centerline{\includegraphics[angle=0,width=17cm]{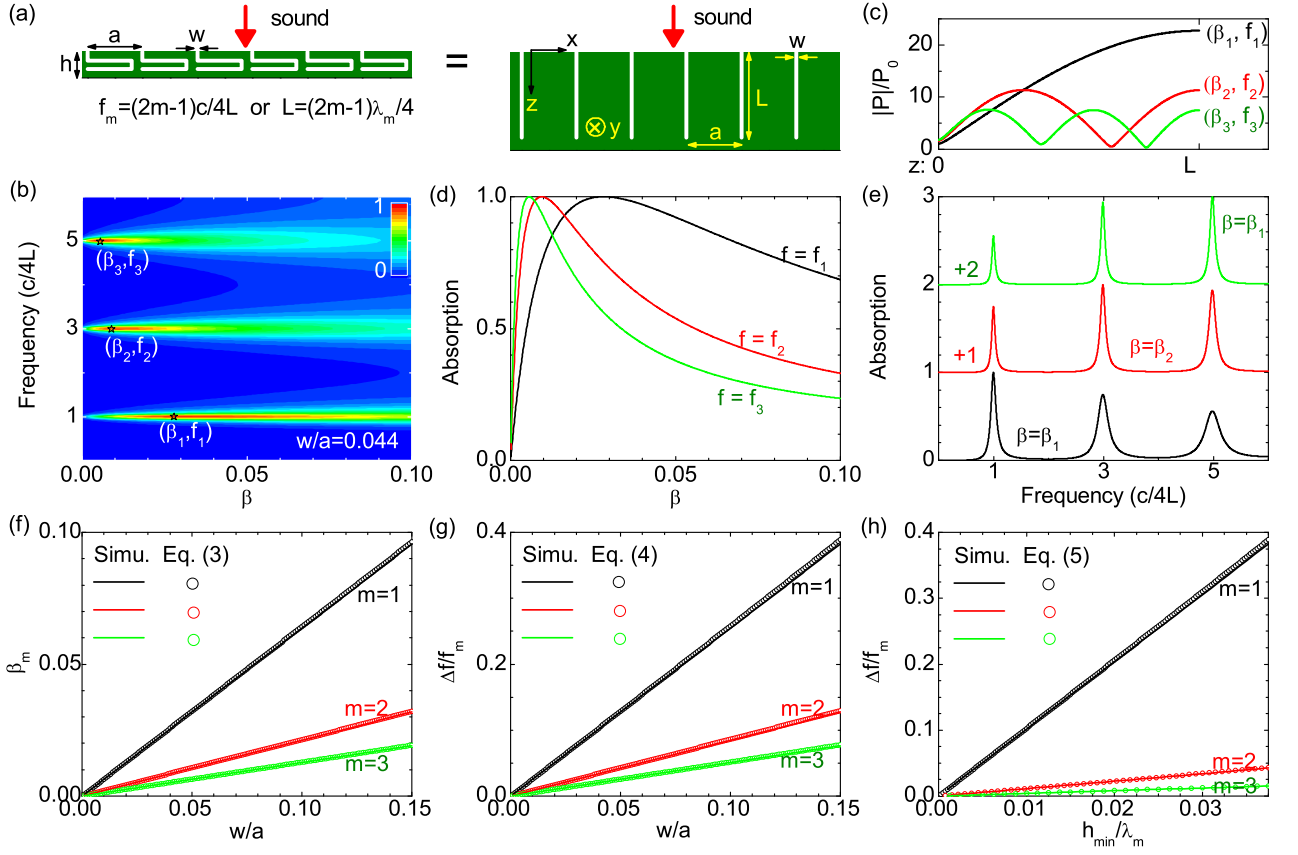}}
\caption{(Color online) Illustrating the design principle of single-port
LAMMs in 2D. (a) A rigid slab with a periodic array of curled,
one-end-closed channels, which is embedded in a background fluid with sound
velocity of $c$ (left panel). The channels have a size $w$, length $L$, and
period $a$. The structure is invariant in the $y$ direction. The slab is
impinged normally by a plane sound wave with wavelength $\protect\lambda $,
frequency $f$, and pressure $P_{0}$. The wavenumber is $k\left( 1+i\protect%
\beta \right) $ in the channels, where $k=2\protect\pi /\protect\lambda $
and $\protect\beta $ represents the loss. The acoustic absorption of the
slab does not change when the channels become straight (right panel). (b)
Calculated absorption, shown in color, as a function of frequency $f$ and
loss $\protect\beta $ for the slab with $w/a=0.044$ in (a). The absorption $%
A=1$ when $f=f_{m}$ and $\protect\beta =\protect\beta _{m}$ with $m=1,2,3$.
(c) The distributions of acoustic pressure inside the channels for the
situations with $A=1$ in (b). (d) and (e) Replotting of (b) along $f=f_{m}$
and $\protect\beta =\protect\beta _{m}$, respectively. (f) and (g) Optimal
loss $\protect\beta _{m}$ and relative bandwidth $\Delta f/f_{m}$ as a
function of channel size $w$, where $\Delta f$ is the full width at half
maximum of the $m$-th absorption peak with $\protect\beta =\protect\beta %
_{m} $. (h) Relative bandwidth $\Delta f/f_{m}$ as a function of $h_{min}/%
\protect\lambda _{m}$, where $\protect\lambda _{m}=c/f_{m}$ and $%
h_{min}=Lw/a $ is the minimal thickness of the slab with curled channels.}
\end{figure*}

\begin{figure*}[tbp]
\centerline{\includegraphics[angle=0,width=17cm]{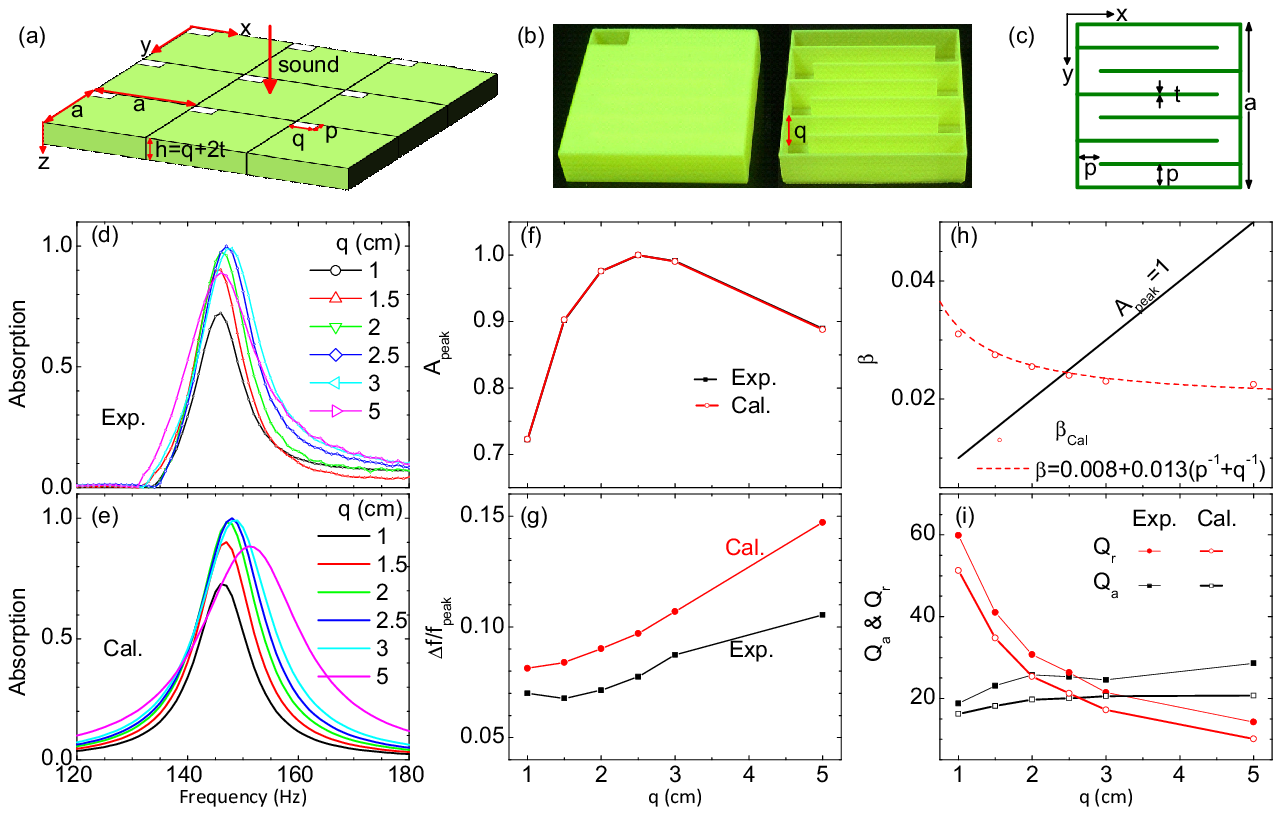}}
\caption{(Color online) Experimental realization of a 3D single-port LAMM
with a narrow absorption band. (a) An acoustic metasurface that is immersed
in air and impinged normally by a plane sound wave. The metasurface is
periodic in the $x$-$y$ plane and its unit cell has a size $a=8.92$ cm in
both the $x$ and $y$ directions and height $h$ in the $z$ direction. A
curled, one-end-closed channel exists in the unit cell and is connected to
the outside via a rectangular aperture at the upper surface of the unit
cell. The aperture has the same area as the cross section of the channel.
(b) Photographs of a realistic unit cell fabricated with PLA by means of 3D
printing (left) and its inner structure (right). (c) Schematic illustration
of the curled channel in the unit cell. The channel has a wall thickness $%
t=1 $ mm, width $p=1.16$ cm and height $q$. (d) and (e) Measured and
calculated absorption spectra for various heights of unit cells in (a).
(f)-(i) Amplitude and relative width of the absorption peak ($A_{peak}$ and $%
\Delta f/f_{peak}$), the loss $\protect\beta $ in channel for achieving
ideal and realistic absorption (black line and symbols in (h)), and the
absorptive and radiative quality factors ($Q_{a}$ and $Q_{r}$) of the unit
cell as a function of the height $h $.}
\end{figure*}

\begin{figure}[th]
\centerline{\includegraphics[angle=0,width=8.5cm]{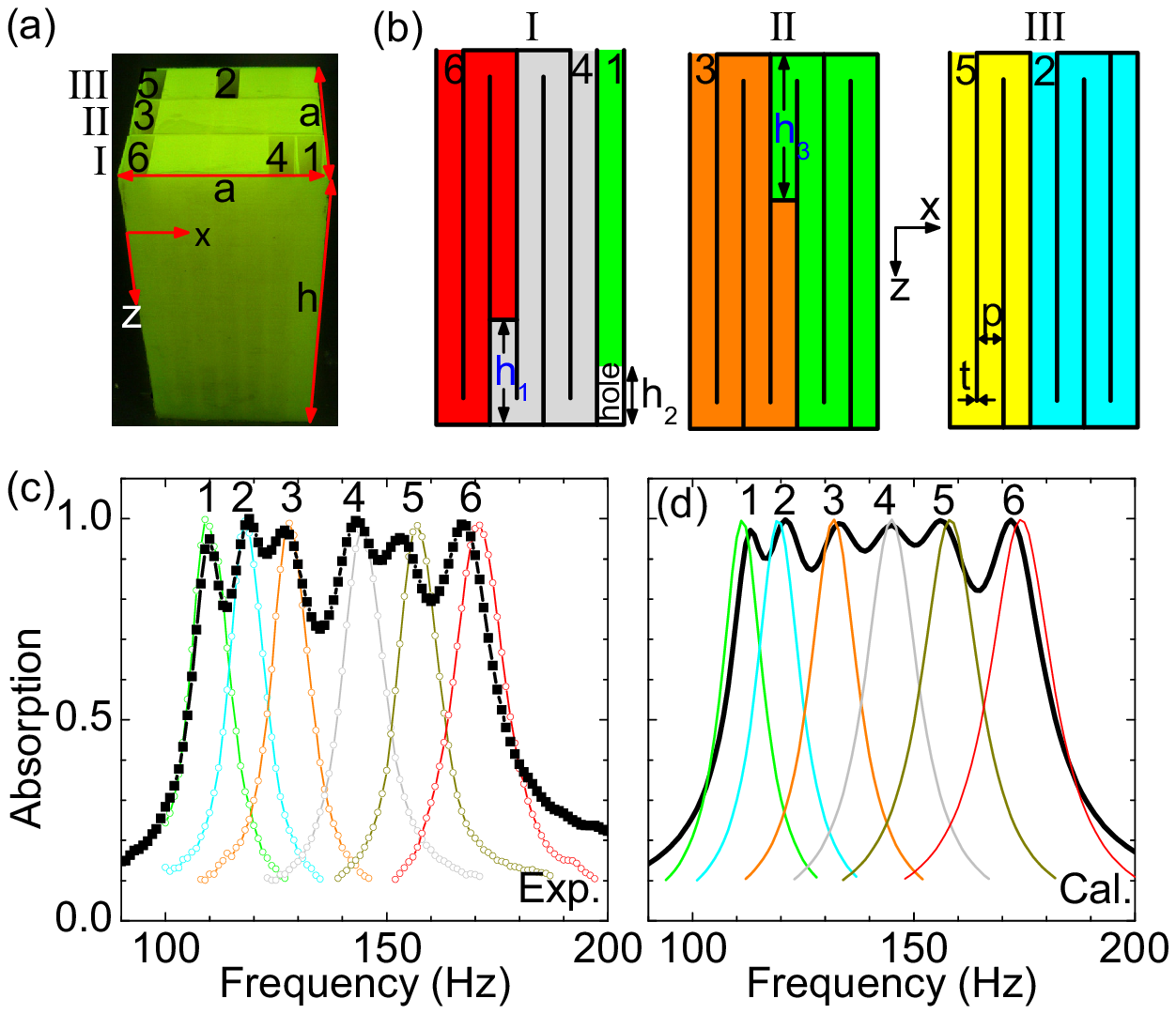}}
\caption{(Color online) Experimental realization of a 3D single-port LAMM
with broadband absorption. (a) Photograph of the unit cell of acoustic
metasurface. The unit cell has a size $a=8.92$ cm in both the $x$ and $y$
directions and height $h=18$ cm, and is composed of three parts that are
fabricated with PLA by means of 3D printing. Six curled, one-end-closed
channels exist in the unit cell and are connected to the outside via
rectangular apertures of $2.84$ cm $\times $ $1.16$ cm at the upper surface
of the unit cell. The lengths of the six channels are 49, 54, 59, 65, 72,
and 79 cm, respectively. (b) Schematic illustration of the six channels
(indicated with different colors) in the unit cell. The channels have a wall
thickness $t=1$ mm and width $p=1.16$ cm. Other parameters are $h_{1}=5$ cm,
$h_{2}=2.84$ cm, and $h_{3}=7$ cm. (c), (d) Measured and calculated
absorption spectra of the unit cell (thick curves). The results are also
plotted when one aperture is open and others are closed in the unit cell
(thin curves).}
\end{figure}

To illustrate the principle, we first consider a 2D-labyrinthine acoustic
metasurface, which is a rigid slab with a channel array embedded [see Fig.
1(a)]. Unlike previous labyrinthine AMMs \cite{Liang1,Xie1,Liang2,Xie2,Cheng}%
, the labyrinthine metasurface here employs one-end-closed channels so that
transmission is forbidden. The metasurface is perpendicular to the $z$
direction, invariant along the $y$ direction, and immersed in a background
fluid where the sound velocity is $c$. The channels have a length of $L$,
width of $w$, and period $a$ along the $x$ direction ($a\gg w$). Upon the
metasurface is normally impinging a harmonic sound plane wave with
wavelength $\lambda \gg a$ and frequency $f=c/\lambda $. Inside the
channels, the wavenumber of sound $k_{c}=k\left( 1+i\beta \right) $, where $%
k=2\pi /\lambda $ and $\beta $ denotes the loss of sound. The system can be
dealt with a coupled mode theory \cite{Supp}. If only fundamental modes are
considered in both the channels and background, the absorption of the
metasurface can be analytically derived \cite{Supp},%
\begin{equation}
A=1-\left\vert r\right\vert ^{2},
\end{equation}%
where the reflection coefficient $r=\left( 1-Z\right) /\left( 1+Z\right) $,
the impedance of the metasurface relative to the background $Z=\left(
w/a\right) \left[ 1-\exp \left( i2k_{c}L\right) \right] /\left[ 1+\exp
\left( i2k_{c}L\right) \right] $, and $i$ is the imaginary unity. Matched
impedance ($Z=1$) and thus unity absorption ($A=1$) can be achieved when
particular values of frequency and loss ($f=f_{m},\beta =\beta _{m}$) are
satisfied. Here, the resonant frequency $f_{m}$ and critical loss $\beta
_{m} $ are given by \cite{Supp}%
\begin{eqnarray}
f_{m} &=&\frac{\left( 2m-1\right) c}{4L}, \\
\beta _{m} &=&\frac{2w}{\pi \left( 2m-1\right) a},
\end{eqnarray}%
where $m$ is a positive integer ($m=1,2,3...$). The corresponding relative
absorption bandwidth can also be analytically obtained%
\begin{equation}
\frac{\Delta f}{f_{m}}=\frac{8w}{\pi \left( 2m-1\right) a}.
\end{equation}%
where $\Delta f$ is the full width of half maximum for the $m$-th order
absorption peak \cite{Supp}. When the rigid body has a minimal filling
fraction in the metasurface, the metasurface has a minimal thickness $%
h_{min}=Lw/a$, so that Eq. (4) can be rewritten as%
\begin{equation}
\frac{\Delta f}{f_{m}}=\frac{32h_{min}}{\pi \left( 2m-1\right) ^{2}\lambda
_{m}}.
\end{equation}%
where $\lambda _{m}=c/f_{m}$ is the wavelength of the $m$-th order
resonance. We note that, if the ratio of $w/a$ is replaced by $pq/a^{2}$
(with $p$, $q$, and $a$ defined in Figs. 2(a)-2(c)), the above formulas are
also valid for a 3D-labyrinthine metasurface. In particular, to achieve
perfect absorption for the fundamental mode ($m=1$) in 3D, the loss in
channels should satisfy%
\begin{equation}
\beta =\frac{2pq}{\pi a^{2}}.
\end{equation}

To verify the accuracy of the above formulas, we did full-wave simulations
for a 2D-labyrinthine metasurface with parameters of $w=0.044a$ and length $%
L=6.7a$. The results are shown in Figs. 1(b)-1(e). We can see that the
metasurface supports multiple resonant modes with frequencies agreeing well
with Eq. (2). For the $m$-th order resonance [see Fig. 1(c)], there are $m$
nodes (with local minimum) of sound pressure along the channel, which are
also antinodes (with local maximum) of fluid velocity. Due to the friction
between the fluid and channel walls, sound waves dissipate inside the
channels. However, only when the sound loss inside channel approaches
critical values ($\beta =\beta _{m}$), the metasurface can absorb totally
sound waves at resonant frequencies ($f=f_{m}$), as shown in Figs. 2(d) and
2(e).

Figs. 1(f)-1(h) demonstrate the critical loss $\beta _{m}$ inside channels
and relative absorption bandwidths $\Delta f/f_{m}$ for different channel
widths. Good agreement is found between the simulated and analytical
results. Both the critical loss $\beta _{m}$ and relative absorption
bandwidths $\Delta f/f_{m}$ are independent of the channel length, and
related only to both the resonance order $m$ and relative channel width $w/a$%
. If a small resonance order (implying few antinodes of fluid velocity) or
large channel width is applied, a high critical loss is needed for achieving
total absorption, leading to a large relative absorption bandwidth.

According to the above theory, we fabricated a 3D-labyrinthine metasurface
which is periodic in the $x$-$y$ plane and embedded in air, as shown in Fig.
2(a). Its building block is a cuboid box fabricated with polylactic acid
(PLA) by means of 3D printing technology [see Fig. 2(b)]. The box has a
fixed size of $a=8.92$ cm in both the $x$ and $y$ direction and a height of $%
h$ in the $z$ direction. A curled channel is embedded inside the box [see
Fig. 2(c)] and connected to outside via an aperture at the upper surface of
the box. Both the opening and cross section of channel are rectangles with
the same width of $p=1.16$ cm and height of $q=h-2t$, where the thickness $%
t=1$ mm for all the walls of the box and channels.

In order to adjust the sound loss in channels, we fabricated a series of
building blocks with the channel height $q$ ranging from 1 cm to 5 cm, and
measured their acoustic absorption spectra [see Fig. 2(d)]. Absorption peaks
can be seen at 146 Hz, corresponding to the $m=1$ resonance of a channel
with length of $L=58$ cm. Such an effective channel length is close to the
realistic length of $7(a-2t)=61$ cm. The channel height has small influence
on the resonant frequency while it strongly affects the amplitude of
absorption peak [see Fig. 2(f)]. When the channel height increases, the peak
amplitude increases first and then decreases. When the channel height $q=2.5$
cm, the measured absorption can be as high as 99.9\% at resonant frequency.
In addition, the relative absorption peak width $\Delta f/f_{peak}$
increases with increasing the channel height (or the metasurface thickness).
For a channel height $q$ of 3 cm, the relative peak width can be 9\%, which
is about ten times of membrane-type metasurfaces \cite{Ma}.

We applied a finite-element method to simulate the 3D-labyrinthine
metasurface \cite{Supp}. The fundamental resonant frequency was first
obtained by using a small sound loss in channels. Then the critical loss in
channels was searched for achieving unity absorption at the resonant
frequency, as shown as the solid line in Fig. 2(h). The simulated results
are very close to the values by Eq. (6). The sound loss in channels was also
obtained [see the symbols in Fig. 2(h)] by fitting the amplitudes of
calculated absorption peaks with the measured ones. We can see that the
sound loss in channels decrease with increasing the channel height $q$. The
obtained sound loss from measured absorption can be fitted by a simple model%
\begin{equation}
\beta =\beta _{0}+2g\left( p+h\right) /ph,
\end{equation}%
where $\beta _{0}=0.005$ is the sound loss in fluid without boundaries, and
the second term with $g=0.08$ is due to the friction between fluid (air)
molecules and channel walls. When the channel height $q=2.5$ cm, the loss in
channels approaches a critical value, leading to unity absorption at
resonant frequency [see Figs. 2(e) and 2(f)]. The relative peak width $%
\Delta f/f_{peak}$ is also found to increase with increasing the channel
height [see Fig. 2(g)].

Apart from the above first-principle microscopic simulation, the
labyrinthine metasurface can also be understood by a macroscopic model,
where the metasurface is regarded as a one-port resonator array with
reflection coefficient given by%
\begin{equation}
r=1-\underset{m}{\sum }\frac{2Q_{r,m}^{-1}}{-i2\left( f/f_{m}-1\right)
+Q_{a,m}^{-1}+Q_{r,m}^{-1}}.
\end{equation}%
Here, $Q_{a,m}=\pi f_{m}t_{a,m}$ and $Q_{r,m}=\pi f_{m}t_{r,m}$ are the
absorptive and radiative quality factors, respectively; $t_{a,m}$ and $%
t_{r,m}$ are the lifetimes of the resonance due to absorption inside the
structure and radiation to the far field, respectively. For the $m=1$ mode,
the absorptive and radiative quality factors, $Q_{a}$ and $Q_{r}$, were
retrieved from the measured and simulated absorption \cite{Supp}, as shown
in Fig. 2(i). We can see that, with increasing the channel height $q$, the
absorptive quality factor $Q_{a}$ increases while the radiative one $Q_{a}$
decreases. When $2$ cm $<q<3.2$ cm, the two quality factors are close to
each other ($Q_{a}\approx Q_{r}$), so that the metasurface can absorb
strongly ($A>98\%$) incident sounds from the background medium of air at
resonant frequency.

The above experiments demonstrate that, when appropriate cross sections of
channels are adopted, a labyrinthine metasurface with deep-subwavelength
thickness ($h/\lambda =1.2\%$) can totally absorb sounds at resonant
frequency of $146$ Hz, holding a relative bandwidth of 9\%. In the
following, we show that the relative absorption bandwidth can be further
broadened by applying various channels in the unit cell of metasurface.

Figure 3(a) shows a fabricated cuboid unit cell with an unchanged size $%
a=8.92$ cm in both the $x$ and $y$ directions and increased height $h=18.2$
cm in the $z$ direction. Inside the unit cell, there exist six channels with
the same cross section ($p=2.84$ cm and $q=1.16$ cm; see Fig. 3(b)). The
channel lengths are $78$, $72$, $66.4$, $59.0$, $54.1$, and $49.7$ cm,
respectively.

Absorption spectra were first measured for unit cells with a single channel
[see Fig. 3(c)], where one channel is hollow and the others were filled with
water. We can see that the six channels individually contribute six
absorption peaks, with central frequencies at 109, 118, 128, 144, 157 and
171 Hz, respectively. Similar to the channel lengths, the six resonant
frequencies are also a geometric progression with common ratio of $1.07$.
All the six absorption peaks have amplitudes higher than 97\% and an average
width about 9\%. When all the six channels are unblocked, the six peaks can
merge together into an absorption band, with frequencies ranging from 105 to
177 Hz and relative width of 51\%. The thickness of the unit cell is much
shorter than the central wavelength of the absorption band ($h/\lambda =0.07$%
).

Although only six channels exist in the unit cell as shown in Fig. 3(a),
more channels can be adopted in the configuration. For an extreme situation
with 21 channels, the first-order absorption band can possess a relative
bandwidth as high as 180\%. In addition, absorption bands with neighbored
orders can overlap with each other, resulting in an ultrabroad absorption
band ($A>95\%$ for $f>f_{c}$) which is similar to that of a porous medium
\cite{Are}. Hence, our 3D single-port LAMMs serve as a bridge linking
damping resonant structures in a narrow band \cite{Ma,Maa} and porous
absorptive media for high frequencies \cite{Are}.

In summary, we have designed, fabricated, and tested a 3D single-port LAMM
which can perfectly absorb airborne sound in a low-frequency band. Both the
position and width of the band can be tuned flexibly. Such a new type of
sound-absorbing materials serves as a bridge between traditional porous
materials for high frequencies and advanced damping resonant structures in a
narrow band. Our work presents a robust approach in controlling the sound
loss in perforations and could benefit experimental realizations of more
acoustic designs based on labyrinthine metamaterials.


\begin{thebibliography}{99}
\bibitem{Liu} Z. Y. Liu, X. Zhang, Y. Mao, Y. Y. Zhu, Z. Yang, C. T. Chan,
P. Sheng, Science \textbf{289}, 1734 (2000).

\bibitem{Li1} J. Li and C. T. Chan, Phys. Rev. E \textbf{70}, 055602 (2004).

\bibitem{Lai} Y. Lai, Y. Wu, P. Sheng, and Z. Q. Zhang, Nature Mater.
\textbf{10}, 620 (2011).

\bibitem{Fang} N. Fang, D. Xi, J. Xu, M. Ambati, W. Srituravanich, C. Sun,
and X. Zhang, Nature Mater. \textbf{5}, 452 (2006).

\bibitem{Hu} X. Hu, K. M. Ho, C. T. Chan, and J. Zi, Phys. Rev. B \textbf{77}%
, 172301 (2008).

\bibitem{Yang1} Z. Yang, J. Mei, M. Yang, N. H. Chan, and P. Sheng, Phys.
Rev. Lett. \textbf{101}, 204301 (2008).

\bibitem{Mei} J. Mei, G. Ma, M. Yang, Z. Yang, W. Wen, and P. Sheng, Nature
Commun. \textbf{3}, 756 (2012).

\bibitem{Yang2} M. Yang, G. Ma, Z. Yang, and P. Sheng, Phys. Rev. Lett.
\textbf{110}, 134301 (2013).

\bibitem{Ma} G. Ma, M. Yang, S. Xiao, Z. Yang and P. Sheng, Nature Mater.
\textbf{13}, 873 (2014).

\bibitem{Li2} J. Li, L. Fok, X. B. Yin, G. Bartal, and X. Zhang, Nature
Mater. \textbf{8}, 931 (2009).

\bibitem{Zhu} J. Zhu, J. Christensen, J. Jung, L. Martin-Moreno, X. Yin, L.
Fok, X. Zhang, and F.J. Garcia-Vidal, Nature Phys. \textbf{7}, 52 (2011).

\bibitem{Park} C. M. Park, J. J. Park, S. H. Lee, Y. M. Seo, C. K. Kim, and
S. H. Lee, Phys. Rev. Lett. \textbf{107}, 194301 (2011).

\bibitem{Lu} M. H. Lu, X. K. Liu, L. Feng, J. Li, C. P. Huang, Y. F. Chen,
Y. Y. Zhu, S. N. Zhu, and N. B. Ming, Phys. Rev. Lett. \textbf{99}, 174301
(2007).

\bibitem{Chr2} J. Christensen, L. Martin-Moreno, and F. J. Garcia-Vidal,
Phys. Rev. Lett. \textbf{101}, 014301 (2008).

\bibitem{Fle} R. Fleury and A. Alu, Phys. Rev. Lett. \textbf{111}, 055501
(2013).

\bibitem{Cum} S. A. Cummer and D. Schurig, New J. Phys. \textbf{9}, 45
(2007).

\bibitem{Chen} H. Chen and C. T. Chan, Appl. Phys. Lett. \textbf{91}, 183518
(2007).

\bibitem{Zhang} S. Zhang, C. Xia, and N. Fang, Phys. Rev. Lett. \textbf{106}%
, 024301 (2011).

\bibitem{Liang1} Z. Liang and J. Li, Phys. Rev. Lett. \textbf{108}, 114301
(2012).

\bibitem{Xie1} Y. Xie, B. I. Popa, L. Zigoneanu, and S. A. Cummer, Phys.
Rev. Lett. \textbf{110}, 175501 (2013).

\bibitem{Liang2} Z. Liang, T. Feng, S. Lok, F. Liu, K. B. Ng, C. H. Chan, J.
Wang, S. Han, S. Lee, and J. Li, Sci. Rep. \textbf{3}, 1614 (2013).

\bibitem{Xie2} Y. Xie, W. Wang, H. Chen, A. Konneker, B.-I. Popa, and S. A.
Cummer, Nature Commun. \textbf{5}, 5553 (2014).

\bibitem{Cheng} Y. Cheng, C. Zhou, B. G. Yuan, D. J. Wu, Q. Wei, and X. J.
Liu, Nature Matter. \textbf{14}, 1013 (2015).

\bibitem{Supp} See the Supplemental Material for derivations of Eqs. (1)-(4), details of simulations,
sample fabrications, and characterizations.

\bibitem{Maa} D-Y. Maa, J. Acoust. Soc. Am. \textbf{104}, 2861 (1998).

\bibitem{Are} J. P. Arenas and M. J. Crocker, J. Sound Vib. \textbf{44},
12-18 (2010).
\end{thebibliography}
\end{document}